\documentclass[useAMS,usenatbib]{mn2e}
\usepackage{graphicx,natbib,times,mathptm,bm, multirow}
\usepackage[fleqn]{amsmath}
\def\Halpha{{\rm H}\alpha}

\bibpunct{(}{)}{;}{a}{}{,}

\begin{document}

\bibliographystyle{mn2e}

\title[Hierarchical Bayesian Extinction Mapping]{3D Extinction Mapping Using Hierarchical Bayesian Models}

\author[Sale]{S.~E.~Sale$^{1,2,3}$\thanks{E-mail: stuart.sale@physics.ox.ac.uk}\\
$^1$ Departamento de F\'{i}sica y Astronom\'{i}a, Facultad de Ciencias, Universidad de Valpara\'{i}so, Av. Gran Breta\~na 1111, Playa Ancha, Casilla 53,
Valpara\'{i}so, Chile\\
$^2$ Departamento de Astronom\'{i}a y Astrof\'{i}sica, Pontificia Universidad Cat\'olica de Chile, Av. Vicu\~na Mackenna 4860, Casilla 306, Santiago 22,
Chile\\
$^3$ Rudolf Peierls Centre for Theoretical Physics, Keble Road, Oxford OX1 3NP, UK\\
}

\date{Received .........., Accepted...........}

\maketitle

\begin{abstract}

	The Galaxy and the stars in it form a hierarchical system, such that the properties of individual stars are influenced by those of the Galaxy. Here, an approach is described which uses hierarchical Bayesian models to simultaneously and empirically determine the mean distance-extinction relationship for a sightline and the properties of stars which populate it. By exploiting the hierarchical nature of the problem, the method described is able to achieve significantly improved precision and accuracy with respect to previous 3D extinction mapping techniques. 

This method is not tied to any individual survey and could be applied to any observations, or combination of observations available. Furthermore, it is extendible and, in addition, could be employed to study Galactic structure as well as factors such as the initial mass function and star formation history in the Galaxy.

\end{abstract}

\section{Introduction}\label{sec_intro}

The Milky Way Galaxy occupies a unique role in astronomy. As we are located within it, we are able to observe and analyse it and its constituents in a manner not possible with any other galaxy. However, this also means that we lack a global view of it. Thus, in order to analyse the Galaxy's structure and history we are forced to infer distances to stars. A task considerably complicated by the requirement to disentangle the effects of interstellar extinction.

Recent years have been characterised by a growth in survey astronomy. Wide area photometric surveys such as 2MASS \citep[][]{Skrutskie_Cutri.2006} and SDSS \citep[][]{York_Adelman.2000} have revolutionised astronomy on all scales, from brown dwarfs to cosmology. Meanwhile, there has also been significant development in surveys specifically studying the Galactic Plane, which is home to the majority of stars in the Galaxy. Such Galactic plane surveys include IPHAS \citep[INT/WFC Photometric $\Halpha$ Survey of the Northern Galactic Plane,][]{Drew_Greimel.2005}, UVEX \citep[The UV-Excess Survey of the Northern Galactic Plane,][]{Groot_Verbeek.2009} and the imminent VPHAS+ in the optical, which will collectively provide narrow and broadband photometry for the entire Galactic disc ($|b|<5$). Meanwhile, in the near infrared UKIDSS--GPS \citep[UKIRT Infrared Deep Sky Survey -- Galactic Plane Survey,][]{Lucas_Hoare.2008} and VVV \citep[Vista Variables in the Via Lactea,][]{Minniti_Lucas.2010} will, between them, cover much of the Galactic plane. The photometric surveys have been accompanied by spectroscopic surveys, including RAVE \citep[][]{Siebert_Williams.2011} and SEGUE \citep[][]{Yanny_Rockosi.2009}.

All these surveys have massively increased the volume of data available on our Galaxy and, with the right analytical tools, should allow us to develop an improved understanding of the Galaxy. This trend for the growth in survey astronomy is set to continue in the next decade; future facilities such as Gaia and LSST will gather data at a hitherto unseen rate. As such, an ongoing problem is how best to harness that data in order to extract as much information as possible \citep[e.g.][]{Binney_only.2011}.

\cite{Juric_Ivezic.2008} used stars from SDSS to study the stellar number density distribution at high Galactic latitudes (mostly $|b|>25^{\circ}$, determining large scale properties of the Galaxy and discovering localised over densities in the Galactic stellar halo. \cite{Ivezic_Sesar.2008} followed this by also analysing the metallicity of stars from SDSS. \cite{Carollo_Beers.2010}, again using SDSS data, claim that they have discovered a second halo component. Though, as argued by \cite{Schonrich_Asplund.2011}, it appears that this second component may be an artefact stemming from a substantial bias present in the distance estimates employed by \cite{Carollo_Beers.2010}.

Studies, such as those discussed in the previous paragraph, demonstrate both the potential rewards and possible pitfalls associated with using large survey datasets. Similarly, it is becoming increasingly clear that simple methods of inference are potentially both ineffective and inaccurate. In large survey datasets various parameters influence data in a manner which is not directly obvious and different parameters will have conflicting effects on the observations. Consider the example of using star counts to estimate the scale length of the Galactic thin disc: a simple minded approach might consist of estimating the distance of the stars by `photometric parallax' and then measuring scale lengths from the resulting distribution. In this case, the individual distance estimates may well be biased and the sample will become incomplete and contaminated in a way which is not known. In short there are a number of confounding factors which prevent the scale length being accurately inferred through simple `inversion'.

Interstellar extinction is a particular hindrance when studying the Galaxy. It builds up gradually, along all lines of sight, in a manner which is not yet well known. A particular realisation of the difficulties interstellar extinction causes is the degeneracy between interstellar reddening and spectral type: an apparently red star may either be an intrinsically red late type star subject to little extinction or a heavily extinguished, intrinsically blue early-type star. Therefore, the presence of extinction makes it difficult to determine the spectral type and thus distance of stars, considerably complicating the study of structure in the Galaxy.

Furthermore, the 3D distribution of extinction is itself intrinsically interesting. Mapping extinction allows one to trace the distribution of dust in the Galaxy. Interstellar dust is itself only one component of the ISM, examining its distribution and comparing it to tracers of other components of the the ISM, such as CO maps \citep[e.g.][]{Dame_Hartmann.2001} or HI maps \citep[e.g.][]{Kalberla_Burton.2005}, offers a window onto the ISM and the physical processes which shape it.

\cite{Juric_Ivezic.2008} were able to correct for the effects of extinction on their observed stars using the asymptotic Galactic extinction map of \cite{Schlegel_Finkbeiner.1998}. This approach was reasonable as the majority of stars in their sample lie at high Galactic latitudes, beyond the Galactic dust layer, which is localised near the Galactic plane.  However, such an approach is not valid for the majority of stars in the Galaxy which lie close to the Galactic plane and are thus unlikely to be beyond all (or nearly all) the Galactic dust in their direction.

In light of the difficulties interstellar extinction causes and the intrinsic value of studying extinction, there has been an ongoing effort to map it over many years, going back to \cite{Trumpler_only.1930} and \cite{vandeKamp_only.1930} noticing the propensity for extinction to be stronger near the Galactic plane. Detailed discussions of the history of extinction mapping are available in \cite{Sale_Drew.2009} and \cite{Majewski_Zasowski.2011}. Recently, spurred on in part by the imminent launch of Gaia and construction of other telescopes, and the availability of high quality data, there has been a renewed focus on extinction mapping \citep{Bailer-Jones_only.2011, Majewski_Zasowski.2011, Schlafly_Finkbeiner.2010, Schlafly_Finkbeiner.2011}.

Most recently \cite{Berry_Ivezic.2011} have used SDSS and 2MASS photometry to estimate the spectral type, distance and extinction towards 73 million stars and so create extinction maps with fine angular scales. In common with many previous efforts to study extinction, they consider stars in isolation, fitting the parameters of a star without reference to the other stars local to it. In doing so they do not take advantage of the fact that properties of stars are correlated. For example, we know that stars located close to each other should be subject to similar extinctions: trivially we would be surprised if two stars closely located on the sky and apparently at similar distances exhibited the effects of vastly differing amounts of extinction. In not employing a method that utilises this information, their results are less precise than what could in principal be achieved with their data.

A contrasting approach is to use Galactic models to study extinction \citep{Marshall_Robin.2006} and Galactic structure \citep[e.g.][]{Robin_Creze.1992, Ruphy_Robin.1996, Sale_Drew.2010}, this offers several distinct advantages. Specifically, analysing stars en masse makes it possible to exploit the physical relationships that exist between stars. Moreover, by comparing Galactic models to real observations, effects such as Malmquist bias, sample contamination and sample incompleteness need not impact upon any inferences, as they will occur in a properly constructed Galactic model in a similar manner to how they occur in reality. However, such approaches typically involve binning the data in some manner. In doing so there is some inevitable loss of information, such that results obtained will be less precise than those obtained without binning. Additionally, such techniques marginalise over the parameters of individual stars, which may be undesirable if these are of interest.

\subsection{Hierarchical Bayesian models}\label{sec_intro_bayes}

Hierarchies pervade almost all aspects of Galactic astronomy. Trivially, the Galaxy is comprised of various stellar (e.g the thin and thick discs) and non stellar (e.g. dark matter halo, ISM) components. These can be broken down into further subcomponents, stars in the stellar case, such that the properties of the stars are dependant on the properties of the stellar component from which they are drawn. Other simple examples of hierarchies include: the distribution of the masses of stars in a newly formed star cluster which are drawn from some initial mass function (IMF) and the kinematics of stars in the Galaxy which are influenced by some global gravitational potential field.

Extinction mapping is a directly hierarchical problem: stars within a field trace a distance--extinction relationship for that sightline. However, existing methods of 3D extinction mapping do not exploit this hierarchical nature. For example, in methods such as those of \cite{Neckel_Klare.1980} and \cite{Arenou_Grenon.1992}, the distance and extinction of each star are first determined, without reference to other stars. Subsequently, the final distance-extinction relationship was estimated by fitting to the determined distance and extinction values for each star. Alternatively, a properly constructed method could use information gained from the field as a whole (i.e distance--extinction relationship) to help constrain the estimates of parameters of individual stars (i.e. their distance and extinction), which in turn can be used to refine our knowledge of the field. Using this form of `group knowledge', exploiting the correlations which exist between stars, enables parameters to be determined more precisely than is otherwise possible. Therefore, \cite{Neckel_Klare.1980} and \cite{Arenou_Grenon.1992}, in common with \cite{Berry_Ivezic.2011}, obtain results less precise than their data are capable of.

Hierarchical Bayesian models are rich statistical models, they extend upon simple Bayesian models by allowing some parameters in the model to be dependant on other parameters. One can, in general, solve for all the parameters in the model, both those at a `lower' level, which may pertain to stars, as well as those at a `higher' level which could describe a sightline or the Galaxy.

On a purely mathematical level, if we have some observation $z$ and wish to estimate some parameter $\theta$ we can employ Bayes' theorem:
\begin{gather}
P(\theta|z)=\frac{P(z|\theta)P(\theta)}{P(z)}\\
P(\theta|z) \propto P(z|\theta)P(\theta)
\end{gather}

\noindent It is also possible to define a hierarchical model, whereby $\theta$ itself depends on a further parameter, often referred to as a hyperparameter, $\phi$. Working from conditional probabilities one can obtain:
\begin{equation}
P(\theta, \phi, z)=P(\theta, \phi|z)P(z)
\end{equation}

\noindent Also:
\begin{align}
P(\theta, \phi, z) &=P(z|\theta, \phi)P(\theta, \phi)\\
&=P(z|\theta, \phi)P(\theta| \phi) P(\phi)
\end{align}

\noindent Therefore:
\begin{equation}
P(\theta, \phi|z) \propto P(z|\theta, \phi)P(\theta|\phi)P(\phi)
\end{equation}

\noindent If $z$ is not directly conditional on $\phi$, this then becomes:
\begin{equation}
P(\theta, \phi|z) \propto P(z|\theta)P(\theta|\phi)P(\phi) \label{eqn:hierarchical_general}
\end{equation}

One could add a further tier to the hierarchy by introducing a further parameter on which $\phi$ depends, this can be repeated as necessary. Also, it is possible to replace the single parameters or observations $z$, $\theta$ and $\phi$ with sets of several parameters. 

It is instructive to consider a simple hierarchical model of two tiers: a top tier containing hyperparameters describing a galaxy and a lower tier of parameters describing the stars which inhabit the galaxy, of which we possess some observations. The form of hierarchical Bayesian model given by equation~\ref{eqn:hierarchical_general} can be employed because any observations of stars we possess are not directly dependant on the state of the Galaxy. The power of the hierarchical model in this instance is clear, it allows information about the galaxy to be estimated directly from the data. That is to say, we can solve for the posterior distribution and from that we possess estimates of certain galaxy-wide parameters that would otherwise be difficult or impossible to determine.

\section{Mapping extinction in 3D}\label{sec_map}

This section describes an algorithm, called H-MEAD (Hierarchically Mapping Extinction Against Distance), which seeks to map extinction along a sightline and determine the properties of stars in this direction. On one level, H-MEAD operates on the idea that extinction can be mapped by following the effect it has on stars. As such, a Hierarchical Bayesian model is constructed which describes the relationship between the observations of stars within a field, the parameters of these stars and the extinction distance relationship which the stars follow. Then, by solving for the posterior distribution it is possible to estimate the parameters of all the stars and the distance--extinction relationship.

\subsection{Without interstellar extinction}\label{sec_map_none}

First let us start with a simple case, considering an individual star in the (assumed) absence of interstellar extinction. Let $\bm{\tilde{y}}_i$ represent the observations we possess for this star. The star has been drawn from a catalogue of stars within some field and to distinguish it from others in the catalogue it is labelled with $i$, the need to do so will become clear in section~\ref{sec_map_unknown}. The observations could be of any form: photometry, astrometry or spectroscopy. However, in this paper only the case of photometric observations will be considered in detail. In the case of multiband photometry, $\bm{\tilde{y}}_i = \{\tilde{y}_i^{(1)}, \tilde{y}_i^{(2)}, \tilde{y}_i^{(3)}, ...\}$, where the $\tilde{y}_i^{(j)}$ represent magnitudes in different bands or repeated observations. For example, in the case of first year VVV data, with one epoch of $ZYJH$ photometry and six epochs in $Ks$, we would have $\bm{\tilde{y}_i} = \{Z, Y, J, H, K_s^{(1)}, K_s^{(2)}, K_s^{(3)}, K_s^{(4)}, K_s^{(5)}, K_s^{(6)}\}$. For IPHAS data, where (almost) every object has been observed twice in its three filters, $\bm{\tilde{y}_i} = \{r'^{(1)}, i'^{(1)}, \Halpha^{(1)}, r'^{(2)}, i'^{(2)}, \Halpha^{(2)}\}$.

$\bm{\tilde{y}}_i$ is an estimate, with errors,  of the true observable parameters of a star, $\bm{y}_i$. Returning to the example of catalogue photometry, $\bm{y}_i$ contains the actual apparent magnitudes of the star. After observing the star we then possess an estimate of the star's apparent magnitudes, with errors, $\bm{\tilde{y}}_i$ in our catalogue, possibly with some observations missing.

One additional factor that must be included in the likelihood is the probability that a star would be included in the catalogue of stars studied. That a star makes it into a sample may be a useful piece of information, depending on how the sample was compiled. Here the event of a given observation of a star being included in the sample or not is given by $^{(j)}S_i$ (where $j$ would iterate over different bands when employing multiband photometry), these are gathered together for each star in the set $\bm{S}_i$.

Then, let $\bm{x}_i$ contain the physical parameters of the star, in which ever way we choose to parametrize them. A simple form would be: $\bm{x}_i=\{{\cal M}_i, \tau_i, [{\rm Fe}/{\rm H}]_i, \mu_i\}$. Where: ${\cal M}_i$ is the star's initial mass, $\tau_i$ its age and $\mu_i$ is the star's distance modulus. Other forms could involve substituting effective temperature and surface gravity for mass and age.

We aim to find the posterior probability distribution of the stellar parameters $\bm{x}_i$, given the observations $\bm{\tilde{y}}_i$. Following from Bayes' theorem:
\begin{equation}
P(\bm{x}_i|\bm{\tilde{y}}_i, \bm{S}_i)=\frac{P(\bm{\tilde{y}}_i, \bm{S}_i|\bm{x}_i)P(\bm{x}_i) }{ P(\bm{\tilde{y}}_i, \bm{S}_i)}
\end{equation}

\noindent As is convention, this equation can simplified, as the evidence, $P(\bm{\tilde{y}}_i, \bm{S}_i)$, is constant for a single model, such that:
\begin{equation}
P(\bm{x}_i|\bm{\tilde{y}}_i, \bm{S}_i) \propto P(\bm{\tilde{y}}_i, \bm{S}_i|\bm{x}_i)P(\bm{x}_i) \label{eqn:posterior_basic}
\end{equation}

\begin{figure}
\centering
\includegraphics[width=80mm]{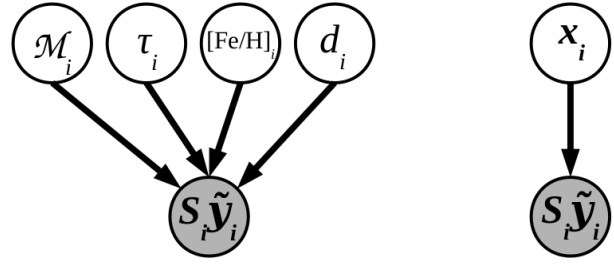}
\caption{A pair of directed acyclic graphs (or Bayesian networks) depicting the dependence of the observable parameters of a star, $\bm{y}_i$ on its physical parameters, $\bm{x}_i$. The two graphs are equivalent, on the left $\bm{x}_i$ is expanded into its component members. These are graphical depictions of the model expressed by equation~\ref{eqn:posterior_basic}. \label{noA_graph}}
\end{figure}

A graphical description of this equation is given in Fig~\ref{noA_graph}, demonstrating how the observations $\bm{\tilde{y}}_i$ depend on the star's physical parameters as contained in $\bm{x}_i$.

\subsubsection{Likelihood}\label{sec_map_none_likelihood}

We first decompose the likelihood as follows:
\begin{equation}
P(\bm{\tilde{y}}_i, \bm{S}_i|\bm{x}_i)=P( \bm{S}_i|\bm{\tilde{y}}_i,\bm{x}_i) P(\bm{\tilde{y}}_i|\bm{x}_i)
\end{equation}

\noindent For many purposes it is sufficient to compile samples of stars from survey photometry, perhaps applying colour or magnitude cuts. As discussed by \cite{Burnett_Binney.2010}, the probability of a star being in such a sample is purely a function of its observed photometry and therefore constant for different $\bm{x}_i$, but identical $\bm{\tilde{y}}_i$. Or more succinctly $\bm{S}_i$ is directly dependant on $\bm{\tilde{y}}_i$ only. As such, it is possible to say:
\begin{equation}
P( \bm{S}_i|\bm{\tilde{y}}_i,\bm{x}_i) = P( \bm{S}_i|\bm{\tilde{y}}_i)
\end{equation}

Continuing to concentrate on the case of multiband photometry, a given set of stellar parameters $\bm{x}_i$ implies that the star $i$ has an actual (not observed) apparent magnitude, $^{(j)}y_i$, in a given band, $j$:
\begin{equation}
^{(j)}y_i(\bm{x}_i)= ^{(j)}M({\cal M}_i, \tau_i, [{\rm Fe}/{\rm H}]_i)+\mu_i
\end{equation}

\noindent Where $^{(j)}M$ is the star's absolute magnitude in that band. Currently, H-MEAD obtains these from the Padova library of isochrones \citep{Marigo_Girardi.2008, Girardi_Williams.2010}, which is, where necessary, converted into the relevant filter system using model spectra \citep[specifically those of][]{Munari_Sordo.2005}.

If it is assumed that the observed apparent magnitudes are normally distributed around their actual value, it is then possible to say that:
\begin{equation}
P(^{(j)}\tilde{y}_i|\bm{x}_i) = \frac{1}{ \sqrt{2 \pi ^{(j)}\delta(\bm{x}_i)^2}} \exp{(\frac{-(^{(j)}\tilde{y}_i - ^{(j)}y_i(\bm{x}_i))^2}{2 ^{(j)}\delta(\bm{x}_i)^2})}
\end{equation}

\noindent Where $^{(j)}\delta(\bm{x}_i)$ is the uncertainty on the apparent magnitude, in the band denoted by $j$, of a star that has the stellar parameters $\bm{x}_i$ and an apparent magnitude $^{(j)} y_i(\bm{x}_i)$. In practice this is currently estimated using a similar technique to \cite{Marshall_Robin.2006} and \cite{Sale_Drew.2010}. In short, $^{(j)}\delta(\bm{x}_i)$ is determined from: 
\begin{equation}
^{(j)}\delta(\bm{x}_i)=A + \exp(B^{(j)} y_i(\bm{x}_i)+C)
\end{equation}

\noindent Where $A$, $B$ and $C$ have been determined by fitting to the estimated photometric uncertainties of objects within the field, produced during pipeline processing of the observations.

Thus, for observations we have obtained:
\begin{equation}
P(^{(j)}\tilde{y}_i, ^{(j)}S_i|\bm{x}_i) = P( ^{(j)}{S}_i| ^{(j)}\tilde{y}_i) P(^{(j)}\tilde{y}_i|\bm{x}_i)
\end{equation}
\noindent As the first term depends only on an observation, this is constant and so:
\begin{equation}
P(^{(j)}\tilde{y}_i, ^{(j)}S_i|\bm{x}_i) \propto \frac{1}{ N(\bm{x}_i) \sqrt{2 \pi ^{(j)}\delta(\bm{x}_i)^2}} \exp{(\frac{-(^{(j)}\tilde{y}_i - ^{(j)}y_i)^2}{2 ^{(j)}\delta(\bm{x}_i)^2})}
\end{equation}

\noindent Where $N(\bm{x}_i)$ is a normalisation factor.

However, in real catalogues of data individual stars may be missing one or more observations which other stars may possess. For example, in a photometric catalogue a star may lack an observation in one band because it was too faint to be observed in that band. In this case $^{(j)}S_i$ will record the absence of an observation. We would like to repeat the process of the previous few paragraphs, but are unable to do so as we do not know what the unobserved apparent magnitude of the star should be. Therefore, we marginalise over the missing observation $^{(j)}\tilde{y}'_i$.
\begin{gather}
P(^{(j)}S_i|\bm{x}_i) = \int P(^{(j)}\tilde{y}'_i, ^{(j)}S_i|\bm{x}_i) \,d^{(j)}\tilde{y}'_i \\
 = \int P(^{(j)}S_i|^{(j)}\tilde{y}'_i) P(^{(j)}\tilde{y}'_i|\bm{x}_i) \,d^{(j)}\tilde{y}'_i \\
 = \int \frac{ P(^{(j)}S_i|^{(j)}\tilde{y}'_i)}{ \sqrt{2 \pi ^{(j)}\delta(\bm{x}_i)^2}} \exp{(\frac{-{(^{(j)}\tilde{y}'_i - ^{(j)}y_i)^2}}{2 ^{(j)}\delta(\bm{x}_i)^2})} \,d^{(j)}\tilde{y}'_i  \label{eqn:completness}
\end{gather}

\noindent In the middle of the catalogue's magnitude range $P(^{(j)}S_i|^{(j)}\tilde{y}'_i)$ will be roughly 0, whilst far outside (very bright or very faint) it will be almost 1. At the bright end it is normally possible to approximate $P(^{(j)}S_i|^{(j)}\tilde{y}'_i)$ as a step function, thus for bright stars equation \ref{eqn:completness} can be approximated with the normal cumulative density function. However, at the faint end, the form of $P(^{(j)}S_i|^{(j)}\tilde{y}'_i)$ is better described with a sigmoid function, the integral in equation~\ref{eqn:completness} must then be solved numerically.

Furthermore, if it is assumed that all members of $\bm{\tilde{y}}_i$ are independent, as would be reasonable for photometry, but not for certain forms of spectroscopically derived measurements based on the same spectrum, the likelihood can be written as:
\begin{equation}
P(\bm{\tilde{y}}_i, \bm{S}_i|\bm{x}_i) \propto \prod_{k} P(^{(k)}\tilde{y}_i|\bm{x}_i) \times \prod_{l} P(^{(l)}S_i|\bm{x}_i)   \label{eqn:likelihood_basic}
\end{equation}

\noindent Where $k$ iterates over observations we have and $l$ those that are missing for each star.

\subsubsection{Prior}\label{sec_map_none_prior}

There are several contributions to the prior. For mass we can take the IMF as a prior, in particular here a Scalo-like IMF is assumed, as it is more suitable for field populations \citep{Kroupa_Weidner.2003}. For age, one could assume the star formation rate, which has been taken to be constant. A radial metallicity gradient can be included in the prior on metallicity, whilst the prior on distance can account both for the radial density gradient of the disc and the geometry of the field (the physical area covered by a source with a finite apparent angular size at a given distance is proportional to the distance squared). As such the prior takes the following form:

\begin{gather}
P({\cal M}_i) \propto {\cal M}_i^{-2.7} \\
P(\tau_i) \propto  {\rm const.}\\
P([{\rm Fe}/{\rm H}]_i) \propto -0.07 R_i \\
P(\mu_i) \propto \rho (\mu_i, l_i, b_i) d_i^3 \\
P(\bm{x}_i)=P({\cal M}_i, \tau_i, [{\rm Fe}/{\rm H}]_i, \mu)=P({\cal M}_i)P(\tau_i)P([{\rm Fe}/{\rm H}]_i)P(\mu_i) \label{eqn:simple_prior}
\end{gather}

\noindent Where $R_i$ is the Galactocentric radius in kpc implied by $\mu_i$ and the Galactic coordinates of the object and $d_i$ is the distance implied by $\mu_i$. 

$\rho (\mu_i, l_i, b_i)$ is the stellar density at a distance modulus of $\mu_i$ in the direction of Galactic coordinates $(l_i,b_i)$. The form of this can be set as desired. H-MEAD, as employed in section~\ref{sec_ver}, models $\rho(\mu_i, l_i, b_i)$ with a thin disc only, as the contribution of other components will be negligible for the cases studied. Following \cite{Sale_Drew.2010} the thin disc is modelled as an exponential truncated disc with an inner scale length of $3000$~pc, a truncation radius of $13$~kpc and an outer scale length of $1200$~pc. Additionally a scale height of $300$~pc is assumed \citep[following][]{Juric_Ivezic.2008}. As discussed in section~\ref{sec_extensions}, a substantially more complicated model for $\rho(\mu_i, l_i, b_i)$, featuring a bulge, thick disc and halo as well as warp and flare in the discs could be employed if desired and it is possible that observations could be used to constrain $\rho(\mu_i, l_i, b_i)$.

Two powers in the $d_i^3$ term in the prior on distance are a result of the fact that all sources subtend non-zero solid angle and that the area covered by this solid angle is proportional to $d_i^2$. The final factor of $d_i$ arises in the Jacobian when converting the prior from one on distance to one on distance modulus.

The Bayesian model described so far is similar that of equation 6 of \cite{Burnett_Binney.2010}. However, there are two key differences: \cite{Burnett_Binney.2010} employ the uncertainties estimated from the observations and then require an extra term to relate these to the uncertainties that would be implied by a given $\bm{x}_i$. Whereas, here uncertainties are estimated directly from $\bm{x}_i$. They also include a separate term which covers the probability of the star being included in the sample, given some $\bm{x}_i$. It should be noted that this is included in the likelihood, thus reconciling their equation 6 and Bayes' equation.

\subsection{Interstellar extinction, with a known distance--extinction relationship}\label{sec_map_known}

The Bayesian model is now extended, by also considering the possibility of interstellar extinction. To do so, $\bm{x}_i$ is altered, such that: $\bm{x}_i=\{{\cal M}_i, \tau_i, [{\rm Fe}/{\rm H}]_i, \mu_i, A_i\}$. 

Note that $A_i$ is monochromatic extinction at a given wavelength. Monochromatic extinction is employed in this study as all broadband measures of extinction are dependant on the SED of the source \citep{McCall_only.2004, Bailer-Jones_only.2011}. As a result, two sources sitting behind the same dust column, but with different SEDs will demonstrate differing $A_V$s (for example), but necessarily identical values of $A_i$. From this point on, the word `extinction' and symbol $A_i$ should both be taken to refer to monochromatic extinction.

\begin{figure}
\centering
\includegraphics[width=80mm]{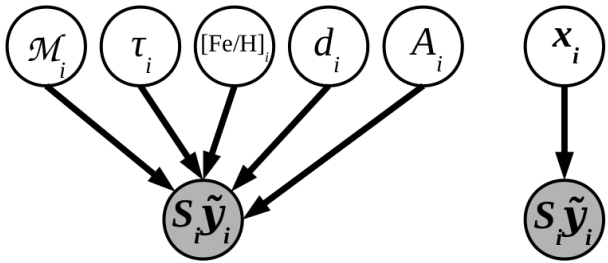}
\caption{Similar to Fig.~\ref{noA_graph}, though now the model includes extinction. \label{knownA_graph}}
\end{figure}

The equation for the posterior probability distribution (equation~\ref{eqn:posterior_basic}) remains the same, though the definition of $\bm{x}_i$ has altered. An updated graphical description is depicted in Fig.~\ref{knownA_graph}.

The definition of the likelihood given by equation~\ref{eqn:likelihood_basic} also remains unchanged. However, the actual apparent magnitude of a star $^{(j)}y_i(\bm{x}_i)$, given stellar parameters is now:
\begin{equation}
^{(j)}y_i(\bm{x}_i)= ^{(j)}M({\cal M}_i, \tau_i, [{\rm Fe}/{\rm H}]_i)+\mu_i+^{(j)}R(\bm{x}_i)A_i
\end{equation}

\noindent Where $^{(j)}R(\bm{x}_i)=^{(j)}A(\bm{x}_i)/A_i$ and is the ratio of broadband extinction in the filter of observation $j$ and the monochromatic extinction for a star with parameters $\bm{x}_i$. Here this ratio is found using the \cite{Fitzpatrick_Massa.2007} $R=3.1$ reddening law and \cite{Munari_Sordo.2005} model spectra.

But what of the prior? Let us say that we know \textit{a priori} the distance-extinction relationship for the field. As all catalogues will inevitably cover some non-zero solid angle, at any given distance there will be a range of possible values of extinction, as the dust column will vary with sky position. This is the effect that causes the well known problem of differential extinction in clusters, where individual stars within a cluster are subject to different extinctions, blurring the appearance of the cluster on colour-magnitude diagrams. We model the differential extinction with a lognormal distribution: at a distance $d$, the distribution of extinction is given by a lognormal distribution with mean $\bar{A}(d)$ and standard deviation $\sigma_A(d)$. \cite{Fischera_Dopita.2004} show, with simulations, that the distribution of extinction to a given distance, with varying angular position, takes a nearly lognormal form, whilst \cite{Goodman_Pineda.2009}, \cite{Froebrich_Rowles.2010} and \cite{Kainulainen_Beuther.2011} have observed that the lognormal form of the column density of ISM components appears to be valid empirically in interstellar clouds. A lognormal distribution can be intuitively seen to be preferable to a Gaussian distribution, as it is only defined for positive values of extinction, i.e. it ensures that there is no probability of the $A_i$ being less than or equal to zero. 

Furthermore, let us describe the distance-extinction relationship $\bar{\bm{ A}} = \{\bar{A}(d), \sigma_A (d)\}$ with piecewise constant function (sometimes called a step or staircase function) for both $\bar{A}(d)$ and  $\sigma_A (d)$.

Therefore, at a given distance, one can say:
\begin{equation}
P(A_i, \mu_i) \propto \frac{1}{A_i \sqrt{2 \pi \upsilon(d_i)^2}} \exp{(\frac{(\log A_i-\zeta(d_i))^2}{2 \upsilon(d_i)^2})} \rho (\mu_i) d_i^2
\end{equation}

\noindent Where $\zeta(d)$ and $\upsilon(d)$ describe a lognormal distribution with mean $\bar{A}(d)$ and standard deviation $\sigma_A (d)$. 

The definition of the overall prior is similar to equation~\ref{eqn:simple_prior}:
\begin{equation}
 P(\bm{x}_i)=P({\cal M}_i, \tau_i, [{\rm Fe}/{\rm H}]_i, \mu_i, A_i)=P({\cal M}_i)P(\tau_i)P([{\rm Fe}/{\rm H}]_i)P(A_i, \mu_i)
\end{equation}

\subsection{The case of an unknown distance-extinction relationship: 3D extinction mapping}\label{sec_map_unknown}

In practice we do not know the distance--extinction relationship \textit{a priori}, in fact we would like to determine it. To this end, we further extend our model to become a hierarchical Bayesian model. Under this description, we allow the parameters contained in $\bar{\bm{ A}}$, which describe the distance-extinction relationship to vary, referring to them as \textit{hyperparameters}.  This makes it possible to constrain the distance-extinction relationship, in addition to the stellar parameters. If only the observations of one star are employed, very little will be learnt of the distance-extinction relationship. However, as the distance--extinction relationship applies to many stars within a field, it can be more precisely determined when using the observations of many stars within the field. Therefore, the tightest constraints on $\bar{\bm{ A}}$ are found when using observations of as many stars as is feasible.

\begin{figure}
\centering
\includegraphics[width=80mm]{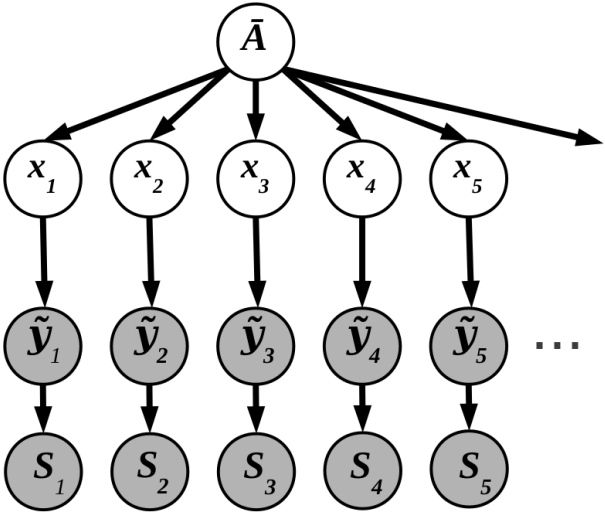}
\caption{A directed acyclic graph depicting of a hierarchical model that combines the distance extinction relationship and stellar parameters. We possess observations $\bm{\tilde{y}}$ of some stars. These observations are dependant on the physical parameters of the stars, $\bm{x}$, which are themselves dependant on the distance--extinction relationship~$\bar{\bm{ A}}$. \label{unknownA_graph}}
\end{figure}

Fig.~\ref{unknownA_graph} shows a graphical description of the hierarchical model that is used in H-MEAD. As can be seen, the stellar parameters, $\bm{x}_i$, of each star in a field are dependant on the distance--extinction relationship that they sample. In practice, this dependence is what allows the distance--extinction relationship to be found.

In the previous sections it has been possible to consider each star separately, as the parameters of one star and those of another were mutually independent. However, now the model includes many stars from within a field. Therefore, for increased clarity, three additional sets are defined, $\bm{x} = \{\bm{x}_1, \bm{x}_2, ...\}$, $\bm{\tilde{y}} = \{\bm{\tilde{y}}_1, \bm{\tilde{y}}_2, ...\}$ and $\bm{S} = \{\bm{S}_1, \bm{S}_2, ...\}$. These sets contain the stellar parameters of all the stars, the observations of all the stars and the set of the events that each star is included in the sample respectively.

Given the model depicted in Fig.~\ref{unknownA_graph} and remembering equation~\ref{eqn:hierarchical_general}, the global posterior distribution is now defined as:
\begin{equation}
P(\bm{x}, \bar{\bm{ A}}|\bm{\tilde{y}}, \bm{S}) \propto P(\bm{\tilde{y}}, \bm{S}|\bm{x})P(\bm{x}|\bar{\bm{ A}})P(\bar{\bm{ A}}) \label{eqn:posterior_hierarchical}
\end{equation}

\noindent The likelihood and prior are each simply the product of the likelihood and prior respectively for all stars:

\begin{gather}
P(\bm{\tilde{y}}, \bm{S}|\bm{x})=\prod_{i} P(\bm{\tilde{y}}_i, \bm{S}_i|\bm{x}_i) \\
P(\bm{x}|\bar{\bm{ A}})=\prod_{i} P(\bm{x}_i|\bar{\bm{ A}})
\end{gather}

It now becomes necessary to set a \textit{hyperprior} -- a prior on the hyperparameters, $P(\bar{\bm{ A}})$. We know that mean extinction ($\bar{A}(d)$) must increase with distance. Therefore, this requirement is included in the hyperprior, by setting the hyperprior probability of any distance--extinction relationships that do not satisfy this requirement to zero.

It was found experimentally that without any further information in the hyperprior, i.e. assuming an otherwise uniform hyperprior, the posterior distribution would not be well behaved. In particular, that $\bar{A}(d)$ would diverge to infinity at large distances. This is a result of a common feature of hierarchical models that, in order for the posterior to be well behaved, the hyperprior should be proper, that is that the integral of the hyperprior across all hyperparameter space should be finite. The hyperprior described in the previous paragraph does not satisfy this requirement.

Therefore, an improved hyperprior was required. As with the priors, the hyperprior is based on existing physical knowledge. Specifically, it is assumed that extinction traces an exponential disc, with scale heights and lengths taken from \cite{Marshall_Robin.2006}. The requirement that mean extinction must increase with distance is also retained in the hyperprior.

\subsection{The hierarchical model in practice}\label{sec_map_practice}

Unfortunately, there exists no analytical solution for the hierarchical model, as given in equation~\ref{eqn:posterior_hierarchical}. Similarly, a `brute force' method, which covers the entire parameter space is clearly unfeasible given the very large number of parameters included in the model. Therefore, a Markov-Chain Monte Carlo (MCMC) method is employed in H-MEAD to estimate the posterior distribution and so, values of each of our parameters.

As a result of the high dimensionality of this problem, a standard Metropolis-Hastings MCMC algorithm \citep{Hastings_only.1970} is not feasible: to maintain a reasonable acceptance rate the proposal distribution would have to be so narrow that it would take an impractically large number of iterations for the chain to converge. Instead a Metropolis-within-Gibbs algorithm \citep{Tierney_only.1994} is adopted. The schema for this is that during each iteration of the chain, each $\bm{x}_i$ is updated in turn, using a Metropolis-Hastings sampler, then the distance--extinction relationship $\bar{\bm{ A}}$ is updated, again using a Metropolis-Hastings sampler. It is worth noting that such `block-updating' algorithms were originally discussed by \cite{Metropolis_Rosenbluth.1953} and therefore such an approach should not be viewed as being particularly radical or exotic.

The feasibility of the  Metropolis-within-Gibbs approach is reliant on the fact that each star's contribution to the posterior probability is dependant only on the $\bm{x}_i$ corresponding to it as well as the distance--extinction relationship. It is independent of the rest of $\bm{x}$, as visualised in Fig~\ref{unknownA_graph}. Thus, when updating a star's parameters only its contribution to the posterior probability need be recomputed. It is only when updating $\bar{\bm{ A}}$ that the contribution of all stars to the posterior need be recomputed and even then it is only necessary to recalculate $P(\bm{x}|\bar{\bm{ A}})P(\bar{\bm{ A}})$. By adopting this method it is also possible to maintain both a reasonable acceptance rate of updates and a sufficiently wide proposal distribution. As a result, the chain quickly converges on the posterior distribution and subsequently samples from it.

The exact performance of any MCMC based algorithm is partly dependant on a number of parameters that describe how it behaves, including the proposal distributions used and the number of iterations the algorithm is allowed to run for. The width of the proposal distributions of the stellar parameters is based on the apparent magnitude of the star: brighter stars being given narrower proposal distributions to reflect the fact that their parameters can be more precisely determined. A `burn in' period of 100000 iterations is assumed; the state of the chain in the first 100000 iterations is not used to estimate the posterior distribution as it is assumed that the state of the chain in these iterations may be affected by the initial values of the parameters. H-MEAD is then allowed to run for a further 150000 iterations, making a total of 250000. The large number of iterations, coupled with the high dimensionality of the parameter space, means it is unfeasible to store the state of the chain at every iteration, instead it is `thinned' by only retaining the state at every 100$^{\rm th}$ iteration.

Up until this point, $\bm{x}_i$ has been defined as $\{{\cal M}_i, \tau_i, [{\rm Fe}/{\rm H}]_i, \mu_i, A_i\}$. However, in practice, it is not possible to use this parameter space as it is extremely difficult for a star's $\bm{x}_i$ to transition from the main sequence to the giant branch: to do so would require the chain to follow a very particular path in $\{{\cal M}_i, \tau_i\}$ space, as such it would take a prohibitively long time for the chain to converge. Therefore, the stellar parameters for each star are parametrized as $\bm{x}_i=\{T^{\rm eff}_i, \log g_i, [{\rm Fe}/{\rm H}]_i, \mu_i, A_i\}$. Everything stated so far in this paper still stands, with the exception that it is necessary to transform the prior probability distribution from $\{{\cal M}_i, \tau_i, [{\rm Fe}/{\rm H}]_i, \mu_i, A_i\}$ space to $\{T^{\rm eff}_i, \log g_i, [{\rm Fe}/{\rm H}]_i, \mu_i, A_i\}$ space, using the Jacobian determinant.

No mention has yet been made of the effect of stellar binarity or higher order multiplicity. Following \cite{Sale_Drew.2009}, a simple correction was made to the estimated distances to all stars such that they are $4\%$ further than would otherwise be calculated.

One could marginalise over $\bm{x}$, so that only the distance--extinction relationship and not any of the stellar parameters are recovered. In practice though there is little to be gained directly from such an approach, marginalising over the stellar parameters in an MCMC algorithm is no quicker than constraining these parameters. A performance increase can only be obtained by binning the data in some form \citep[as in][]{Marshall_Robin.2006}, which induces a loss of precision.

\section{Observations}\label{sec_obs}

This paper concentrates on the use of multiband photometry. Particular reference has been made to the use of observations from IPHAS and VVV, two surveys of the Galactic Plane, where the need for extinction mapping is most acute. It is, though, obvious that the method could be employed with any photometric survey. The hierarchical model employed remains the same, but suitable isochrones must be obtained and the response to extinction in all filters computed.

\begin{figure}
\centering
\includegraphics[width=\columnwidth]{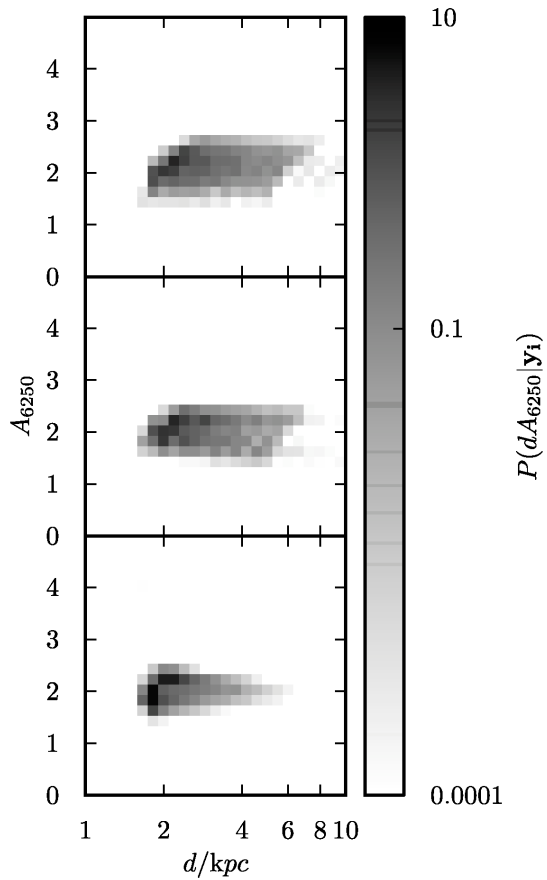}
\caption{The posterior probability distributions of distance and extinction for an A0 star at $2~kpc$ in three different filter systems. Top VISTA, middle SDSS and bottom IPHAS. \label{A_info}}
\end{figure}

\begin{table*}
\caption{ A summary of the histograms in Figs.~\ref{A_info}, \ref{A_info_ext} and~\ref{A_info_parallax}. Columns show the mean of the posterior distribution, and a 68\% credible interval centred on the mean. \label{Adist_ests}}
\begin{tabular}{c || c | c | c | c | c | c | c | c | c | c | c | c}
\multirow{3}{*}{Data} & \multicolumn{4}{|c|}{Without Gaia, uninformative prior on $A,d$} & \multicolumn{4}{|c|}{Realistic prior on $A,d$} & \multicolumn{4}{|c|}{With Gaia}\\
& \multicolumn{2}{|c|}{Mean} & \multicolumn{2}{|c|}{CI} & \multicolumn{2}{|c|}{Mean} & \multicolumn{2}{|c|}{CI} & \multicolumn{2}{|c|}{Mean} & \multicolumn{2}{|c|}{CI}\\
& $d$ / pc & $A_{6250}$ & $d$ / pc & $A_{6250}$ & $d$ / pc & $A_{6250}$ & $d$ / pc & $A_{6250}$ & $d$ / pc & $A_{6250}$ & $d$ / pc & $A_{6250}$ \\ 
\hline
VISTA & 2700 & 2.1 & 870 & 0.5 & 2300 & 2.0 & 220 & 0.3 & 2000 & 2.0 & 60 & 0.1\\
SDSS & 2500 & 2.0 & 650 & 0.5 & 2200 & 2.0 & 190 & 0.2 & 1950 & 2.0 & 50 & 0.1\\
IPHAS & 2300 & 2.0 & 240 & 0.3 & 2100 & 2.0 & 120 & 0.2 & 2030 & 2.0 & 50 & 0.1\\
\end{tabular}
\end{table*}

It is, however, important to note that not all observations are equally informative. To demonstrate this photometry was simulated for an A0 star on the ZAMS, located at a distance of 2~kpc, in the VISTA, IPHAS and SDSS filter systems. This simulated star is assumed to lie towards the anticentre and has been subjected to an extinction of $A_{6250}=2$, that is an extinction equivalent to 2~mag at $6250~\AA$, assuming the \cite{Fitzpatrick_Massa.2007} $R=3.1$ extinction law. This simulated star was then analysed using the method of section~\ref{sec_map_known}. In the first instance a uniform and therefore uninformative prior on extinction was assumed. This is a simple case and is analogous to extinction mapping methods which do not use the distance-extinction relationship to refine the estimates of stellar parameters. The subsequent posterior probability distributions for the distance and extinction of the stars, marginalised over all other variables, are shown in Fig.~\ref{A_info} and summarised in table~\ref{Adist_ests}.

Although all three systems obtain reasonable estimates of the star's distance and extinction, the IPHAS data allows the posterior to be significantly more constrained and thus, demonstrably, carries more information. The particular cause for this is due to the design of the filter sets, the SDSS and VISTA systems are more affected by what is often referred to as a degeneracy between estimated spectral type (and thus equivalently estimates of $T_{\rm eff}$ or mass) and estimated extinction. Due to the nature of the IPHAS filter system, which allows $(r'-\Halpha)$ to be a rough proxy for spectral type, this degeneracy is less dramatic in the estimate derived from IPHAS photometry.

In this case, it would appear that all three systems produce a biased estimate of distance. However, this arises as only one star is being considered and it has been chosen to be on the ZAMS and will therefore be intrinsically faint for a star of its colour. It is of course possible to have an intrinsically brighter star of the same colour, say one that is turning off the main sequence. Such stars will be less common, but in order for them to have the same apparent magnitude as the hypothetical A0 ZAMS star at 2~kpc, they will have to be more distant. The possibility of the observation being of such a star elongates the posterior distribution. In effect, the estimate of the star's absolute magnitude or $\log g$ is not well constrained. When a whole population of stars is considered, on average this effect disappears as the real distribution of these stars will resemble the posterior.

\begin{figure}
\centering
\includegraphics[width=\columnwidth]{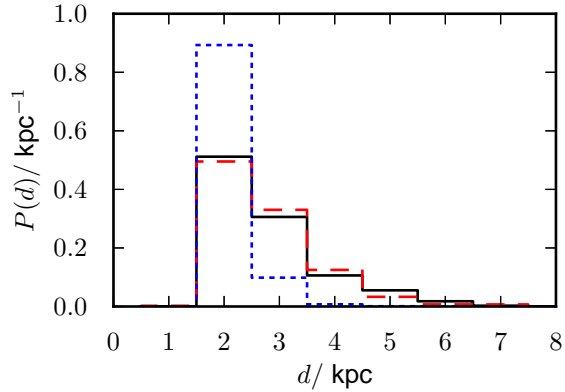}
\caption{A comparison between the posterior distribution for VVV data in Fig.~\ref{A_info} (black solid line) and the distribution of the true distances of simulated stars with apparent magnitudes consistent with that of the A0 star (red dashed line). The posterior distribution for IPHAS data is depicted with the blue dotted line for comparison.\label{A_dist_comp}}
\end{figure}

To verify this we turn to a simulation of VVV photometry, of the type discussed in section~\ref{sec_ver}, though with extinctions of all amounts assumed equally likely, was performed for a region of 200 square degrees. A sample was then populated with all the stars with apparent magnitudes in all three bands within 0.02 that of the A0-star. Fig.~\ref{A_dist_comp} shows a normalised histogram of the true distance of the stars in this sample compared to the posterior distribution for the A-star obtained using simulated VVV data marginalised over all parameters except distance. Fig.~\ref{A_dist_comp} reveals that the posteriors in Fig.~\ref{A_info} are not biased, but are an honest reflection of what can be determined about the star using the data at hand.

\begin{figure}
\centering
\includegraphics[width=\columnwidth]{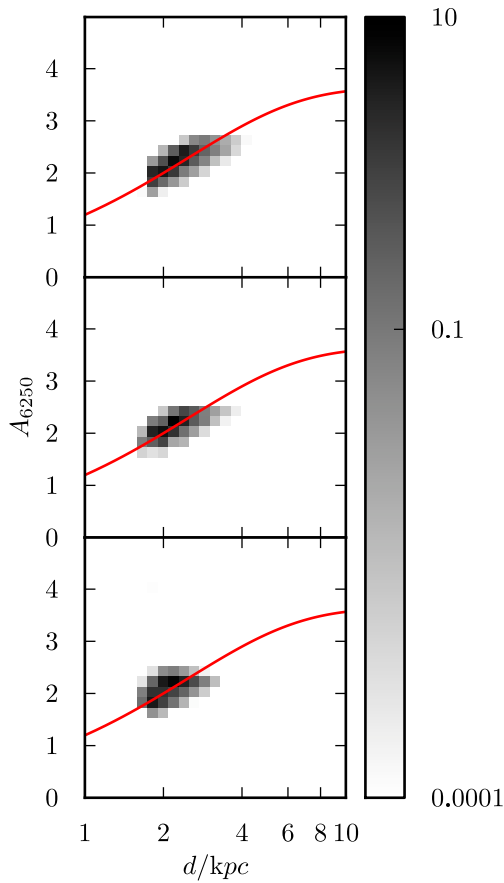}
\caption{Similar to Fig.~\ref{A_info}, though now the depicted (red line) distance-extinction relationship has been employed to set the prior probability. Differential extinction has been set by assuming $\sigma_A(d) = \bar{A}(d)/5$ \label{A_info_ext}}
\end{figure}

\begin{figure}
\centering
\includegraphics[width=\columnwidth]{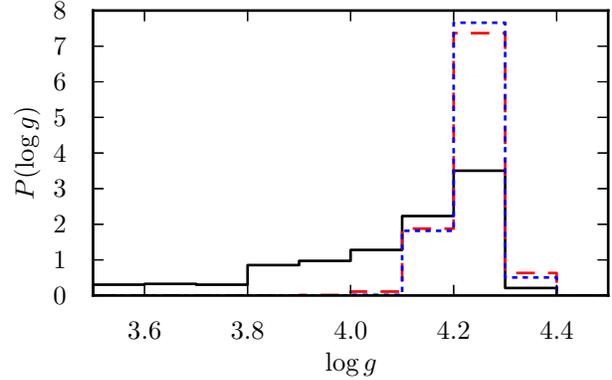}
\caption{A comparison between the marginalised posterior distributions on $\log g$ determined with an uninformative uniform prior (solid black line) and a more informative prior (dashed red line) with SDSS data, and with combined SDSS and Gaia data (blue dots). The surface gravity of the simulated star is $\log g =4.28$ \label{logg_ext}}
\end{figure}

H-MEAD, however, does not assume a uniform prior on extinction, but rather, as indicated by equation~\ref{eqn:posterior_hierarchical}, uses the estimated distance-extinction relationship. To illustrate the effect of assuming a more informative prior on extinction, the analysis that produced Fig.~\ref{A_info} was repeated, though now with a typical distance-extinction relationship used to set the prior. The posteriors and the assumed distance-extinction relationship are depicted in Fig.~\ref{A_info_ext} and the posteriors summarised in table~\ref{Adist_ests}. The resultant priors are significantly more precise, demonstrating the value of the hierarchical approach pursued by H-MEAD. In particular, the extra information from the prior now allows $\log g$ to be more accurately determined, as can be seen in Fig.~\ref{logg_ext}.

The method described here can be applied to almost any form of astronomical observations. Although this work largely concentrates on the use of photometry, it is possible to also use spectroscopy or spectroscopically derived measurements, for example one could use the spectra of SEGUE or RAVE and/or the estimated stellar parameters derived from them. Though this comes with the caveat, as mentioned in section~\ref{sec_map_none_likelihood}, that care must be taken with the likelihood when multiple measurements, for example those produced by the Segue Stellar Parameter Pipeline \citep[SSPP][]{Lee_Beers.2008}, are derived from the same spectra. Additionally, unlike photometric surveys which blindly observe all visible objects within a field of view, targets for spectroscopic surveys are preselected, with preference being given to certain types of object. Therefore, it is necessary to understand this selection function and include it in the likelihood.

\begin{figure}
\centering
\includegraphics[width=\columnwidth]{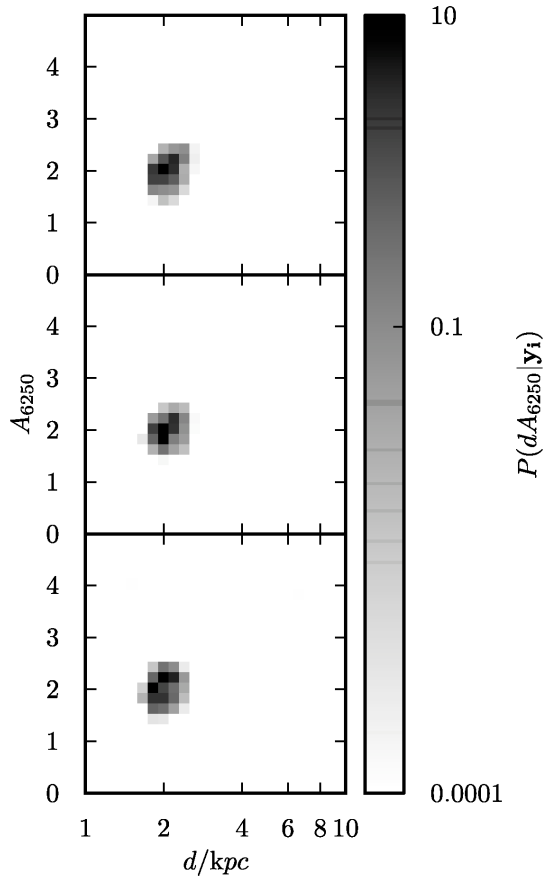}
\caption{Similar to Fig.~\ref{A_info}. However, now a simulated Gaia parallax has been employed in addition to the photometry. \label{A_info_parallax}}
\end{figure}

It is also possible to use astrometric observations, parallaxes being of particular interest in this context. Though they may appear to only constrain stellar distances directly, as estimates of other parameters will be correlated with the estimated distance, they too will be constrained by parallax measurements. As an example, Fig~\ref{A_info_parallax} returns to the example of an A0 star at 2~kpc, subject to an extinction of $A_{6250}=2$. The estimated uncertainty on the Gaia parallax has been derived following \cite{Lindegren_only.2009}. The inclusion of a Gaia like parallax not only allows the distance estimate to be much more tightly constrained, but also improves the precision of the extinction estimate substantially. Furthermore, as the distance is more tightly constrained, so too are estimates of the star's absolute magnitude and $\log g$.

Clearly, the ideal approach would be to use all data available and so obtain the most precise results. In practice though, there is a cost in both CPU time and human effort associated with the use of any individual survey dataset, stemming from the need to configure isochrones and the response to extinction and process the observations in H-MEAD. A sensible approach would be to prioritise the use of more informative surveys over those which will be of less use.

\section{Verification}\label{sec_ver}

It is necessary to assess the veracity of the H-MEAD algorithm, to measure its precision and demonstrate its accuracy. This has been done through the use of simulated photometry, as it is possible to know the exact distance-extinction relationship and the stellar parameters of the stars in the simulation. This process is similar to that described in \cite{Sale_Drew.2009}. The Galactic model employed is largely identical to that used in \cite{Sale_Drew.2010}, which itself follows in the spirit of the Besan\c{c}on model. In broad terms, the model seeks to simulate photometry for an observer in a galaxy similar to the Milky Way. The model includes features such as photometric errors, binarity, metallicity variation and extinction.

\begin{figure}
\centering
\includegraphics[width=\columnwidth]{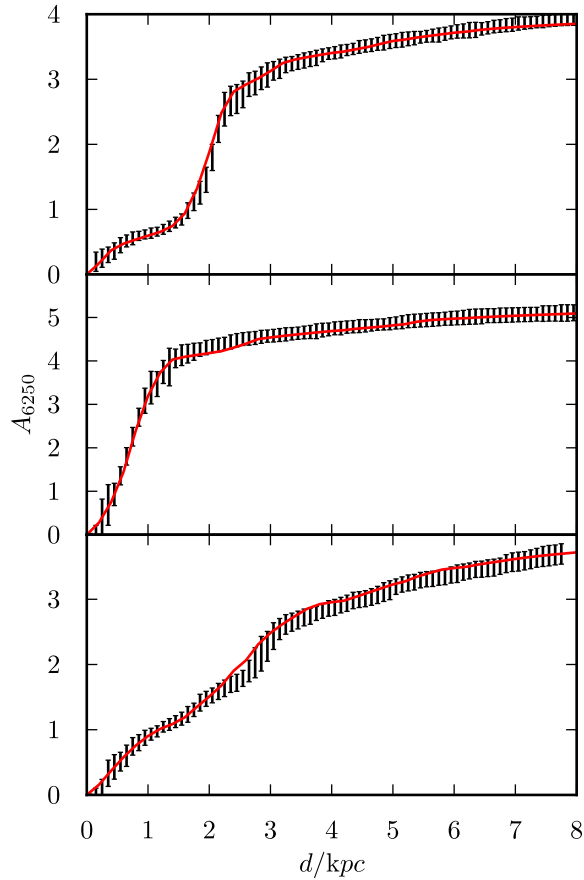}
\caption{The solid red lines show the extinction distance relationships assumed in the production of synthetic photometry. The black error bars show the extinction distance relationship obtained by H-MEAD when run on the synthetic photometry. The sightlines shown are, from top to bottom: $(l,b)=(90,2), (135,2), (180, 0)$. The number of stars in the simulated catalogues are 1766 for $(90,2)$, 1246 for $(135,2)$ and 977 for $(180,0)$. \label{no_m_no_b}}
\end{figure}

As a first test IPHAS photometry was simulated for stars using a Galactic model that features no stellar binarity and a fixed metallicity at solar levels. Subsequently, as in \cite{Sale_Drew.2009}, colour cuts were applied to the simulated data to exclude stars with colours consistent with having a spectral type later than K4. A number of Galactic plane sightlines were examined ($l= 45, 90, 135, 180$ and $b=0, 2, 5$) with the expectation that these should be broadly representative of the different regimes encountered in the Galactic plane. All simulated sightlines had an angular size of $ 5 \arcmin \times 5 \arcmin$, a bright photometric limit at $r'=13$ and a faint limit at $r'\sim 20.5$. In all cases H-MEAD successfully retrieved the input extinction distance relationships, examples can be seen in Fig~\ref{no_m_no_b}.

In Fig.~\ref{no_m_no_b} and subsequent similar figures, error bars are employed to depict $68\%$ credible intervals on the mean extinction in each distance bin. A credible interval contains a given proportion of the posterior probability distribution on one parameter, marginalising over all other parameters, and can be viewed as being an approximate Bayesian equivalent to a confidence interval. In this instance the credible intervals have been estimated from the chains created by the MCMC process. Credible intervals, like confidence intervals, could be placed anywhere in the parameter's range as long as they contain the desired proportion of the posterior probability. In H-MEAD and this paper, for the sake of simplicity and consistency, all credible intervals are centred on the mean value and are symmetric. Note that these error bars do not show differential extinction, $\sigma_A (d)$.

\begin{figure}
\centering
\includegraphics[width=80mm]{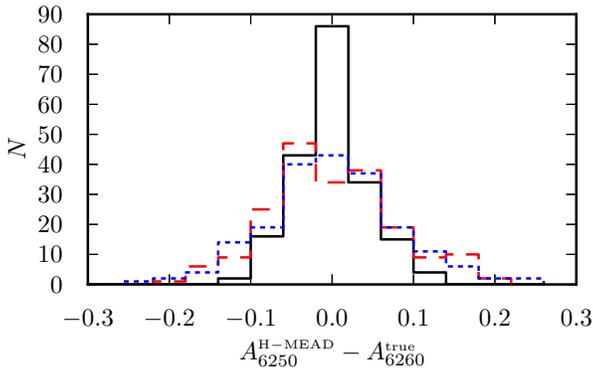}
\caption{The estimated and true values of $\bar{A}(4 {\rm kpc})$ (solid black line), $\bar{A}(5 {\rm kpc})$ (dashed red line) and $\bar{A}(6 {\rm kpc})$ (short-dashed blue line) for 200 different runs of H-MEAD, using different photometric catalogues. The typical half-width of a 68\% credible interval on $\bar{A}$ is $\sim 0.05$ at 4~kpc and $\sim 0.08$ at 5 and 6~kpc. \label{post_sum}}
\end{figure}

To verify that the means and credible intervals are an adequate summary of the posterior on mean extinction, the results of repeated use of H-MEAD were examined. To this end, 200 different synthetic photometric catalogues were simulated for the same sightline, $(l,b)=(180,0)$, H-MEAD was subsequently run on each catalogue and the results analysed. Fig~\ref{post_sum} shows the distribution of the means of the posterior on mean extinction at three different distances and demonstrates that the means and credible interval give a reasonable description of the posterior.

One key feature of the extinction curves in Fig.~\ref{no_m_no_b} that should be noted is that values of mean extinction in successive bins are highly correlated. As a result, it may appear that the credible intervals for mean extinction, displayed as error bars, are underestimated in regions. For example at a distance of $\sim 2.5~kpc$ along the $(180, 0)$ sightline shown in Fig.~\ref{no_m_no_b}, where the estimated extinction is lower than the true value. However, in reality, as the mean value of extinction in successive bins is so strongly correlated the probability of successive bins lying roughly one credible interval below the actual value is only slightly less than that of one bin being similarly placed.

\begin{figure}
\centering
\includegraphics[width=80mm]{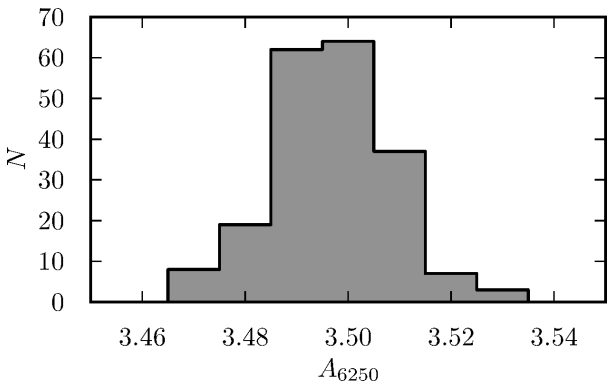}
\caption{The estimated value of $\bar{A}(5 {\rm kpc})$ for 200 different runs of H-MEAD, using the same simulated photometric catalogue. The standard deviation of this distribution is 0.009. For comparison, the 68\% credible interval on $\bar{A}(5 {\rm kpc})$ in each of the 200 runs is typically 0.08. \label{MCMC_error}}
\end{figure}

Using the simulated photometry is was also possible to verify that, in all cases, H-MEAD was able to converge on a result quickly, typically within the first 10,000 iterations; allowing for a burn in period of 100,000 iterations was found to be somewhat conservative. As MCMC algorithms sample from the posterior distribution, quantities derived from the MCMC chain are subject to a standard error. In addition samples drawn from an MCMC chain are not independent, thus the effective size of such a sample is smaller than its absolute size. As a result of the high dimensionality of the statistical model employed in H-MEAD, the MCMC chains mix somewhat slowly. This, in turn, reduces the effective sample size of the chain and thus increases the standard error on the estimates of the distance extinction relationship and stellar parameters. For this reason, the long chains are necessary. In the tests conducted effective sample sizes were typically a few hundred. In comparison, the thinned post-burn-in chain contains 1500 iterations and is itself obtained from the unthinned post-burn in chain of 150000 iterations. Fig.~\ref{MCMC_error} demonstrates that the standard error on the estimated parameters is acceptably small compared to the width of the posterior distribution, thus it can be neglected.

\begin{figure}
\centering
\includegraphics[width=80mm]{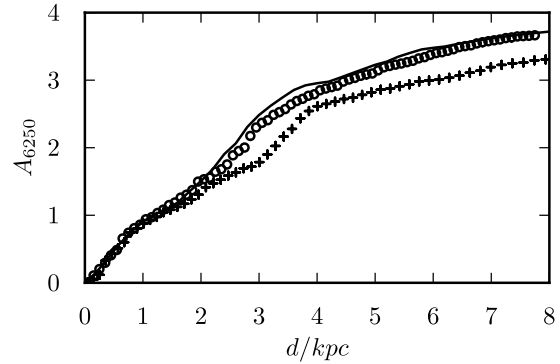}
\caption{Demonstrating the effect of neglecting metallicity variation in stars. The solid line shows the distance--extinction relationship employed to simulate photometry. The crosses show the distance--extinction relationship obtained by H-MEAD, if it assumes all stars have solar metallicity, whilst the circles show that obtained if metallicities are allowed to vary. \label{m_no_b}}
\end{figure}

Colours in the IPHAS system are largely independent of metallicity, magnitudes though are somewhat affected by metallicity variations \citep{Sale_Drew.2009}. As such, the distances to stars estimated with IPHAS photometry are dependant on the contents of the prior. If photometry with colours sensitive to metallicity, e.g. the $(u'-g')$ colour in SDSS \citep{Ivezic_Sesar.2008}, were used instead the metallicity of the stars would be tightly constrained by the data and less dependant on the metallicity gradient in the thin disc assumed in the prior.

The importance of the assumed thin disc metallicity gradient is demonstrated in Fig.~\ref{m_no_b}. IPHAS photometry was simulated, assuming a metallicity gradient of $-0.07$~kpc$^{-1}$ \citep[following][]{Robin_Reyle.2003}. If H-MEAD assumes this same metallicity distribution in the prior, it accurately retrieves the distance--extinction relationship. If, however, it assumes that all stars are of solar metallicity, it overestimates distances significantly. Smaller changes impact on the results far less significantly, for example assuming a flattening of the metallicity gradient at large Galactocentric radii \citep[following e.g.][]{Carraro_Geisler.2007, Bragaglia_Sestito.2008}.

H-MEAD also estimates the stellar parameters for every star in the sample. The precision with which it is able to do so is inevitably dependant on the apparent magnitude of stars. However, it is also affected by the characteristics of the sightline: it is possible to estimate the distance of a star in a region densely filled with dust, where extinction is building up rapidly, more accurately than that of a star in a region where the dust content is low. 

\begin{figure}
\centering
\includegraphics[width=\columnwidth]{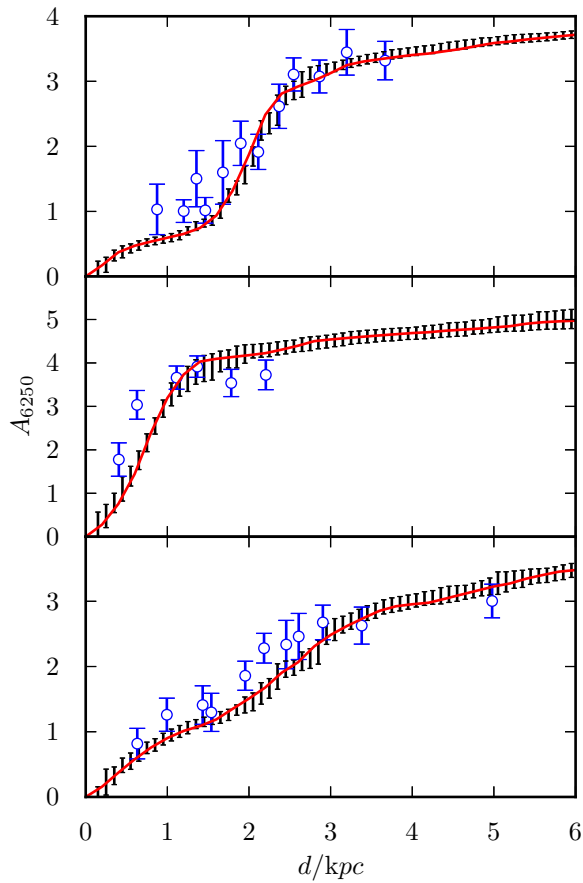}
\caption{Comparing the performance of MEAD and H-MEAD. Again, the sightlines shown are, from top to bottom: $(l,b)=(90,2), (135,2), (180, 0)$. The number of stars in the simulated catalogues are 1692 for $(90,2)$, 1257 for $(135,2)$ and 952 for $(180,0)$. The results obtained by H-MEAD are shown with black error bars and those of MEAD with broader blue bars, with a circle at the mean value.\label{MEAD_compare}}
\end{figure}

Finally a direct comparison is made between the performance of H-MEAD and MEAD \citep{Sale_Drew.2009}. Both algorithms were run on the same simulated photometry, some examples are shown in Fig.~\ref{MEAD_compare}. In all cases H-MEAD produces clearly far superior results.

H-MEAD has already been used with real data. Raddi et al. (in prep.) applied H-MEAD to IPHAS photometry along lines of sight towards 68 Classical Be stars. Armed with the produced distance-extinction relationships and spectroscopically estimated reddenings, `extinction distances' \citep[e.g.][]{Giammanco_Sale.2011} are estimated for each star. These are then compared to distances estimated purely from photometry and so the comparison between the two constitutes a strong test of the capabilities of H-MEAD.

\section{Further extensions to the model}\label{sec_extensions}

In the hierarchical model employed by H-MEAD, as specified by equation~\ref{eqn:posterior_hierarchical} and Fig.~\ref{unknownA_graph} there are a number of fixed parameters. These are found in the prior probability distribution and include the shape of the IMF, the stellar density distribution, the Galactic SFH and how metallicity varies with position in the Galaxy. Poor choice of these priors will lead to systematic error in the estimates obtained by H-MEAD.

Although it has not yet been done, it is in theory possible to further extend the model such that all these parameters are themselves allowed to vary and thus we gain additional hyperparameters. As an example, we could allow the exponent in the assumed IMF to vary. Extending our model in this way would, in the worst case scenario, allow for marginalisation over different IMFs. However, if the data to hand are suitably informative, it would be possible to more tightly constrain the shape of the IMF. By performing this form of analysis, the possibility of systematic error will be much reduced.

In the case of stellar density, one could employ a more complicated model. The prior employed in section~\ref{sec_ver} only includes a simple exponential thin disc when describing stellar densities. This is a good model at low Galactic latitudes, but will be increasingly insufficient with increasing distance from the Galactic plane. One could therefore employ a model which also includes thick disc, bulge and halo components. Additionally, it would also be possible to include warping and flaring of the thin disc in the model. Again the parameters that describe the warp and flare could be allowed to vary and so, the observational data could constrain the form of the thin disc's warp and flare.

The model currently employed assumes an $R=3.1$ reddening law of \cite{Fitzpatrick_only.2004}, following \cite{Cardelli_Clayton.1989}. Clearly, this is a somewhat simplistic approach, given that it is well known that the extinction law varies across the sky. An alternative approach might be to allow the extinction law to vary between fields. However, this too can been seen to be simplistic: if the extinction law is allowed to vary between adjacent fields, why shouldn't it vary within a field? A far more comprehensive would be to model $R$ in the same way as extinction is modelled in this paper. Namely, that within a field, at a given distance, $R$ has some distribution with non-zero width and this distribution is allowed to vary with distance. Again this approach requires a hierarchical model, each star has its own value of $R$, which samples the distribution of $R$ along the line of sight. 

The value of this approach can be most easily seen with a simple example: Imagine that within some field the first few kpc are dominated by `normal' $R=3.1$ dust, then at a distance of a few kpc there exists a dark cloud, containing high $R$ dust covering part of the field. Behind the dark cloud $R=3.1$ dust is again dominant. Assuming $R=3.1$ would clearly be unsatisfactory for stars behind the dark cloud, whilst taking some field averaged value of $R$ would not be representative of either those stars obscured by the cloud or those not. In this case allowing $R$ to have a distribution with non-zero width and allowing this distribution to change with distance would capture the existence of the cloud at some depth along the line of sight and cope with the variation of extinction law with angular position, caused by the cloud not covering the entire field.

In addition, by not treating stars individually, it is possible to take advantage of the fact the stars located near each other should exhibit correlated values of $R$ as they lie behind similar dust columns. As a result, the estimated values of $R$ for each star and for the field in general will be more precise than if the stars were treated individually, in much the same way as the hierarchical approach allows extinction to be measured more precisely.

\begin{figure}
\centering
\includegraphics[width=80mm]{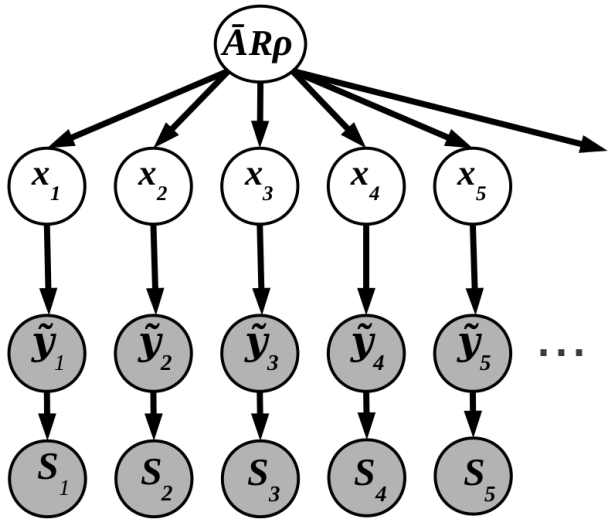}
\caption{A graphical description of an extended hierarchical model which can also be used to study Galactic stellar density~($\bm{\rho}$) and 3D reddening law variation~($\bar{\bm{ R}}$). \label{extended}}
\end{figure}

An example of a model which could be employed to study Galactic structure as well as map extinction in three dimensions is shown in Fig.~\ref{extended}. This model is somewhat more complicated than that employed by H-MEAD and as such will require superior data to constrain the additional parameters. Additionally, it is likely that its implementation will present a number of additional practical hurdles, akin to those discussed in section~\ref{sec_map_practice}.

\section{Closing discussion}\label{sec_close}

Given the high frequency of hierarchies in Galactic astronomy, hierarchical Bayesian models will be a key tool for analysing the ever growing quantity of data on our Galaxy. The advantage of using hierarchical models is clear. Such models relate the parameters of many stars, through the characteristics of the Galaxy or the region in which they are found. Therefore, they provide additional information and allow the parameters which describe stars and those which describe the Galaxy to be determined more precisely and accurately. Conversely, methods which consider stars individually do not take advantage of the relationship between stars and the Galaxy they populate and cannot take full advantage of the data they employ.

Furthermore, employing a Bayesian methodology offers several advantages over classical frequentist techniques. Bayesian techniques avoid the classic inverse problem, whereby the parameters of interest are determined directly from the data, by instead finding the model which best describes the data. Such inversion can be particularly difficult when there are many parameters involved and their estimates are correlated. Additionally, with Bayesian techniques one is able to include prior information, in this case the basic physics of extinction and the accumulated historical knowledge of the Galaxy can both be included to give more precise and accurate results.

This paper has examined one potential use of hierarchical Bayesian models, employing them to map extinction in three dimensions. A good knowledge of where and how extinction builds up in the Galaxy is a major barrier to furthering our understanding of the structure an history of the Galaxy. There are few existing 3D extinction maps and those that do exist are limited by one or more of several factors, including the data they employ \citep[e.g.][]{Neckel_Klare.1980} a reliance on a particular Galactic model \citep{Marshall_Robin.2006}, by only being able to achieve a coarse distance resolution \citep[e.g.][]{Majewski_Zasowski.2011} or by only treating individual stars in isolation \citep[e.g.][]{Berry_Ivezic.2011}.

A particular algorithm, H-MEAD, which employs hierarchical Bayesian models has been described in detail. The model employed has been discussed, concentrating on the physical justification for its use, but also covering some of the practical aspects of its implementation. Subsequently, testing with synthetic photometry has demonstrated that H-MEAD does indeed produce results which are both accurate and precise. The precision of H-MEAD is massively improved with respect to MEAD \citep{Sale_Drew.2009}, whilst the fact that H-MEAD is able to cope with features such as variation in stellar densities and metallicities will considerably reduce systematic errors.

It should also be remembered that H-MEAD not only maps extinction, but also provides precise estimates of the distances, masses, metallicities, etc of stars. There are many potential uses for such data, one possibility is studying the distribution of high mass stars in order to map the Galaxy's spiral structure.

In the longer term, it is not possible to consider the future usefulness of the method described in this paper and hierarchical models in general without mentioning future surveys. Gaia is expected to provide parallaxes for $\sim10^9$ stars as well as spectroscopy for a subset. As such, it will clearly be central to any future analysis of the Galaxy. However, it will be far from the only data available. Rather than simply using Gaia data alone, a more precise analysis will be possible if data from existing and forthcoming surveys, be they photometric or spectroscopic, were used as well. 

However, a major problem is this wealth of data should be analysed. Clearly, concealed within the data will be a gold-mine of information on our Galaxy and, in theory, it should prove possible to make spectacular advances in understanding the current and past state of the Galaxy. This massive advantage though becomes a major hurdle, in so far as the sheer quantity and complexity of the data will considerably complicate its analysis.

The method discussed in this paper, could be considered to be a prototypical example of how Gaia data might best be analysed. There are many potential improvements to the method, several of which have been discussed in section~\ref{sec_extensions}, which will not only increase its accuracy but also extend the scope of what it can achieve. If applied this would make it possible to not only map extinction in 3D, but also map Galactic structure, estimate the Galaxy's star formation history and more besides. 

\section*{Acknowledgements}

The author would like to thank P.~Newcombe and J.\,E.~Drew for discussions on the subject of this paper. The author would also like to thank an anonymous referee, whose thoughtful comments improved the paper.

Support for the author is provided by the Ministry for the Economy, Development, and Tourism's Programa Iniciativa Cient\'{i}fica Milenio through grant P07-021-F, awarded to The Milky Way Millennium Nucleus.

\bibliography{astroph_3,bibliography-2}

\end{document}